# A Multi-State Model for the Reliability Assessment of a Distributed Generation System via Universal Generating Function

Yanfu Li[1], Enrico Zio[1,2]

[1] *Ecole Centrale Paris - Supelec, Paris, France,*

yanfu.li@ecp.fr, yanfu.li@supelec.fr, enrico.zio@ecp.fr, enrico.zio@supelec.fr

[2] *Politecnico di Milano, Milan, Italy, Dipartimento di Energia*

Enrico.zio@polimi.it

**Abstract**

The current and future developments of electric power systems are pushing the boundaries of reliability assessment to consider distribution networks with renewable generators. Given the stochastic features of these elements, most modeling approaches rely on Monte Carlo simulation. The computational costs associated to the simulation approach force to treating mostly small-sized systems, i.e. with a limited number of lumped components of a given renewable technology (e.g. wind or solar, etc.) whose behavior is described by a binary state, working or failed. In this paper, we propose an analytical multi-state modeling approach for the reliability assessment of distributed generation (DG). The approach allows looking to a number of diverse energy generation technologies distributed on the system. Multiple states are used to describe the randomness in the generation units, due to the stochastic nature of the generation sources and of the mechanical degradation/failure behavior of the generation systems. The universal generating function (UGF) technique is used for the individual component multi-state modeling. A multiplication-type composition operator is introduced to combine the UGFs for the mechanical degradation and renewable generation source states into the UGF of the renewable generator power output. The overall multi-state DG system UGF is then constructed and classical reliability indices (e.g. loss of load expectation (LOLE), expected energy not supplied (EENS)) are computed from the DG system generation and load UGFs. An application of the model is shown on a DG system adapted from the IEEE 34 nodes distribution test feeder.

*Key Words: Distributed generation, reliability assessment, multi-state modeling, universal generating function*

**Notations**

| | |
|---|---|
| $s$ | Solar irradiance |
| $n_{IR}$ | Total number of discretized solar irradiance states |
| $s_i$ | Discretized solar irradiance at state $i$ |
| $G^{IR}$ | Random variable representing the power output of one solar module |
| $u_{IR}(z)$ | $u$-function of the power output of one solar module, $G^{IR}$ |
| $n_{sm}$ | Number of functioning solar modules in the solar generator |
| $g_i^{IR}$ | Power output of a single solar module at solar irradiance state $i$ |
| $p_i^{IR}$ | Probability of solar irradiance being at state $i$ |
| $n_{MA}$ | Total number of mechanical states of the renewable generator (e.g. solar generator, wind turbine, and electrical vehicle aggregation) |
| $g_i^{MA}$ | State value of the mechanical state $i$ of the renewable generator |
| $p_i^{MA}$ | Probability of the renewable generator being at mechanical state $i$ |
| $G^{MA}$ | Random variable representing the mechanical condition of the renewable generator |
| $u_{MA}^{(\cdot)}(z)$ | $u$-function of the mechanical condition of one renewable generator, $G^{MA}$ |
| $G^S$ | Random variable representing the power output of a solar generator |
| $g_i^S$ | Power output of a solar generator at performance state $i$ |
| $p_i^S$ | Probability of the solar generator being at performance state $i$ |
| $u_S(z)$ | $u$-function of the solar generator power output, $G^S$ |
| $P_r^S$ | Rated power output of one solar generator |
| $\otimes_\times$ | Multiplication operator of u-functions |
| $v$ | Wind speed |
| $n_{WS}$ | Total number of discretized wind speed states |
| $v_i$ | Discretized wind speed at state $i$ |
| $G^{WS}$ | Random variable representing the power output of one wind turbine at different wind speed levels |
| $g_i^{WS}$ | Power output of a wind turbine at wind speed state $i$ |
| $p_i^{WS}$ | Probability of the wind speed being at state $i$ |
| $u_{WS}(z)$ | $u$-function of the power output of wind turbine at different wind speed levels, $G^{WS}$ |
| $G^W$ | Random variable representing the power output of a wind generator |
| $g_i^W$ | Power output of a wind generator at performance state $i$ |
| $p_i^W$ | Probability of a wind generator being at performance state $i$ |
| $u_W(z)$ | $u$-function of the wind turbine power output, $G^W$ |
| $P_r^W$ | Rated power output of the wind turbine |
| $G^{OP}$ | Random variable representing operation status of one electric vehicle (EV) |
| $g_i^{OP}$ | Power output of an EV at operation state $i$ |
| $p_i^{OP}$ | Probability of an EV being at operation state $i$ |
| $u_{OP}(z)$ | $u$-function of the power output of an EV, $G^{OP}$ |
| $G^{EV}$ | Random variable representing the power output of an EV aggregation |
| $g_i^{EV}$ | Power output of an EV aggregation at performance state $i$ |
| $p_i^{EV}$ | Probability of an EV aggregation being at performance state $i$ |
| $u_{EV}(z)$ | $u$-function of the power output of one EV aggregation, $G^{EV}$ |
| $g_i^T$ | Power output of a transformer being at performance state $i$ |
| $p_i^T$ | Probability of a transformer being at performance state $i$ |
| $u_T(z)$ | $u$-function of the power output of one transformer |
| $g_i^G$ | System power generation at state $i$ |
| $p_i^G$ | Probability of the system power generation being at state $i$ |

$u_G(z)$    *u*-function of the system power generation
$g_i^L$    Power load demand at state *i*
$p_i^L$    Probability of the power load demand being at state *i*
$u_L(z)$    *u*-function of the power load demand
$\Psi(\cdot)$    distributive operator of u-functions

# 1 Introduction

Traditionally, the power distribution network was designed to carry electricity from the transmission devices and delivers it to the end consumers without any energy generation. The reliability assessment of these systems (Allan 1994, Billinton and Allan, 1996) aims at evaluating the sufficiency of the generation facilities output to satisfy the consumer demand (i.e. power delivered exceeding load power consumption). In the past decades, renewable energy sources have become increasingly present in the power distribution network due to the rising prices of conventional energy sources and the enhanced public concerns on environmental issues such as global warming. While the renewable energy sources are increasingly regarded as cost-effective, their power outputs are largely dependent on external natural resources such as solar irradiation and wind speed. Owing to the random nature of these resources, the renewable generators behave quite differently from the conventional generators. This has put additional pressure on the need to assess the reliability of the power distribution network with renewable energy sources (Hegazy et al. 2003).

Monte Carlo simulation is the mainstream method for computing the reliability indices of various sub-areas of a power network such as the generation system (Billinton and Bagen 2006, Billinton and Gao 2008), the transmission network (Papakammenos and Dialynas 2004), the composite generation and transmission system (Gao et al. 2009) and the distribution network (Hegazy et al. 2003, Atwa et al. 2010). In general, simulation can be very effective to approximate the power system behaviors but it can suffer from unstable accuracy and lengthy computation time, which may lead to consider simplifying assumptions, e.g. binary-state representations of the working/failure processes (Hegazy et al. 2003, El-Khattam et al. 2006). Analytical models (Billinton and Allan 1996, Billinton and Gao, 2008, Ding et al. 2011) are also used for reliability assessment, e.g. by Contingency Enumeration or State Enumeration (Beshir et al. 1996). They are usually preferred on small scale power networks (such as local distribution networks), because they can provide more precise estimation of the power adequacy with relatively smaller computation efforts than simulation models (Rei and Schilling, 2008).

For more realistic reliability assessment of power systems, multi-state models, which have been widely applied to resolve system reliability assessment problems (Lisnianski et al., 2010, Natvig, 2011), are being introduced to describe the random behaviors of the generation sources (Billinton and Gao, 2008) and the degradation/repair of the components (Massim et al. 2006). Compared to binary-state models, the multi-state models offer greater flexibility in the description of system state evolution, for more accurate approximation of the real-world power systems (Ding et al. 2006, Zio et al. 2007).

In this paper, we consider a relatively complete electrical network with distributed generation and describe it as a multi-state stochastic system by means of an analytical probabilistic model. To improve the descriptive power and solution efficiency of analytical probabilistic models of multi-state stochastic systems, the Universal Generating Function (UGF) technique has been introduced (Ushakov 1987). It is a powerful analytical tool to describe multi-state components and construct the overall model of complex multi-state systems. It has a wide range of successful applications in various fields (Lisnianski and Levitin, 2003). In the power field, it has been used in particular to model wind generators (Wang et al. 2009, Ding et al. 2011). In our work, the UGF technique is used to model the power output of different random generation sources (Billinton and Gao, 2008) and the degradation/failure/repair behavior of the components of the generation system. The universal generating functions (here after termed u-functions) modeling a generic generation source and the degradation/failure/repair behavior of the generation system components are combined by a multiplication operator formally defined to obtain the u-function of the renewable generator power output. The u-functions modeling the different generation units which make up the distributed generation network (solar generators, wind turbines, electrical vehicles and transformer), are combined to obtain the model of the overall distributed generation system, which is then solved to calculate the reliability indices by taking into consideration the uncertain load demand, also represented by a u-function.

The rest of the paper is organized as follows. In Section 2, the multi-state models of the components (solar generator, wind turbine, electric vehicle, transformer and loads) are developed and mathematically presented by way of the UGF technique. In Section 3, the overall multi-state (stochastic) system (MSS) model is built. The reliability indices are introduced and formulated in terms of UGFs. Section 4 presents a case study and the results obtained with the proposed analytical model. Section 5 provides some discussions, and possible future extensions of the study.

## 2. Multi-State Models of the Individual Components in the Distributed Generation System

This Section describes the multi-state models of the components in the DG system and uses UGF for the mathematical representation.

### 2.1 Solar Generator

We consider a photovoltaic (PV) solar generator made of a number of cells. The model consists of two parts: the solar irradiation function and the power generation function which links the solar irradiation to the power output of the PV. In literature, the Beta distribution has been used to represent the random behavior of the solar irradiation (Ettoumi et al. 2002, Atwa et al. 2010):

$$f(s) = \begin{cases} \frac{\Gamma(\alpha+\beta)}{\Gamma(\alpha)\Gamma(\beta)} \cdot s^{(\alpha-1)} \cdot (1-s)^{\beta-1} & for\ 0 \leq s \leq 1, \alpha \geq 0, \beta \geq 0 \\ 0 & otherwise \end{cases} \quad (1)$$

where $s$ is the solar irradiance kW/m$^2$, $f(s)$ is the Beta function of $s$, $\alpha$ and $\beta$ are the parameters of the Beta function which can be inferred from estimates of the mean and variance values of historical irradiance data. Within a multi-state modeling framework, the continuous solar irradiation distribution needs to be transformed into a discrete distribution (Kaplan 1981, Atwa et al. 2010). To this aim, $s$ is divided into $n_{IR}$ states of equal size, the probability of the $i$th state being:

$$Pr(s_i) = \int_{(i-1)\cdot \Delta s}^{i\cdot \Delta s} f(s)ds \quad (2)$$

where $\Delta s = s_{max}/n_{IR}$ is the step size, and $s_i$ is the value of solar irradiance in the $i$th state:

$$s_i = \frac{i\cdot \Delta s + (i-1)\cdot \Delta s}{2} \quad (3)$$

Once the irradiation distribution is modeled, the output of one solar generator can be determined by the following power generation function (Mohamed and Koivo 2010):

$$P(s_i, n_{sm}) = n_{sm} \cdot FF \cdot V_y \cdot I_y$$

$$I_y = s_i \cdot [I_{SC} + k_i(T_c - 25)]$$

$$V_y = V_{oc} - k_v \cdot T_c$$

$$T_c = T_a + s_i \cdot \frac{N_{ot} - 20}{0.8}$$

$$FF = \frac{V_{MPP} \cdot I_{MPP}}{V_{oc} \cdot I_{sc}} \qquad (4)$$

where $P(s_i, n_{sm})$ is the output power of the solar generator at irradiance level $s_i$ with $n_{sm}$ functioning solar modules, $k_v$ is the voltage temperature coefficient V/°C, $k_i$ is the current temperature coefficient A/°C, $FF$ is the fill factor which is defined as the ratio of the actual maximum obtainable power to the theoretical (not actually obtainable) power, $I_{sc}$ is the short circuit current in A, $V_{oc}$ is the open-circuit voltage in V, $I_{MPP}$ is the current at maximum power point in A, $V_{MPP}$ is the voltage at maximum power point in V, $N_{ot}$ is the nominal operating temperature in °C, $T_c$ is the cell temperature in °C, $T_a$ is the ambient temperature in °C. Given the relation between $s_i$ and $P(s_i)$, the probability of the power state $P(s_i)$ is $Pr(s_i)$.

In solar generation, there are two different sources of randomness: one is the external solar irradiation (Moharil and Kulkarni, 2010), and the other is the internal mechanical degradation of the hardware elements (e.g. the failure of a solar module) and the repairs (Billinton and Karki, 2003). We assume that they are independent from each other. For solar irradiation, in the left portion of Fig. 1 state '0' represents the amount of solar irradiation that transforms into no power generation, '$n_{IR}$-1' represents the state of solar irradiation which produces the maximum power output given that all solar modules are working. For the internal mechanical degradation/failure/repair states (Fig. 1, right), for simplicity we assume that each solar module has only two states (working or failed), leading to a total of $n_{MA}$ states of the solar generator (from complete failure when none of its modules are working to perfect working when all its modules function). In the right portion of Fig. 1, state '0' represents failure of all modules of the solar generator, '$n_{MA}$-1' represents its perfect functioning when all solar modules are producing; the intermediate states represent partial failures, i.e. the loss of a portion of the modules (Goel and Gupta 1993). The solar modules are considered connected in parallel within the structural logic of the solar generator, but physically their failures reduce the capacity of operation percentually.

The random solar irradiation and the degradation/failure/repair process jointly determine the overall power output of the solar generator. This is modeled by eq. (4), which combines by multiplication the number of functioning solar modules (representing the generator mechanical condition) with the power output of the individual solar modules (determined by the strength of solar irradiance). For example, when 50% of the solar modules are working and the solar irradiance is such that each solar module produces 50% of the rated power output, the overall power output of the entire solar generator is 25% of the rated power. The modeling assumption is that the rated power of the solar generator is produced under the perfect condition that all solar modules are working and the solar irradiation is at its peak value.

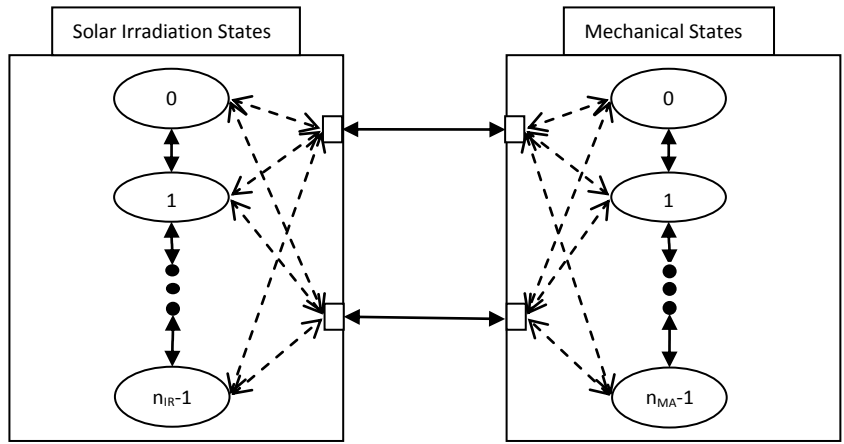

Fig 1. Solar irradiation states and mechanical states of the solar generators

Let $G^{IR}$ and $G^{MA}$ be the discrete random variables representing the states of solar irradiance and mechanical condition, respectively, and let $\overrightarrow{Pr_{IR}} = \{p_0^{IR}, p_1^{IR}, \ldots, p_{n_{IR}-1}^{IR}\}$ and $\overrightarrow{Pr_{MA}} = \{p_0^{MA}, p_1^{MA}, \ldots, p_{n_{MA}-1}^{MA}\}$ denote the state probability distributions of $G^{IR}$ and $G^{MA}$, respectively. The UGF approach is utilized to represent the probability mass functions (PMFs) of the two random variables and derive the PMF of the solar generator power output, $G^S = P(s_i, n_{sm})$. For the solar irradiance, the u-function links the probability of state $i$, $p_i^{IR}$, to the corresponding state value of $G^{IR}$, denoted as $g_i^{IR}$ (i.e. the power output of a single solar module at solar irradiance level $i$):

$$u_{IR}(z) = \sum_{i=0}^{n_{IR}-1} p_i^{IR} z^{g_i^{IR}} \qquad (5)$$

Similarly, the u-function of the mechanical condition is defined as:

$$u_{MA}^S(z) = \sum_{i=0}^{n_{MA}-1} p_i^{MA} Z^{g_i^{MA}} \tag{6}$$

where $g_i^{MA}$ denotes the state value of $G^{MA}$ (in the case of the solar generator, it is the number of working solar modules).

In the theory of UGF, the composition operator $\otimes_f$ is used to derive the u-function of the output variable of an arbitrary structure/property function $f(G_1, G_2, \ldots G_n)$ of $n$ independent random variables $G_1, G_2, \ldots G_n$ (Levitin 2005). For the solar generator model, we introduce a multiplication-type structure function of the two random variables to describe the power output coherently with the mathematical representation of eq. (4). For two random variables $G_1$ and $G_2$, the purposely defined composition operator $\otimes_\times$ then needs:

$$u(z) = u_1(z) \otimes_\times u_2(z) = \sum_{i=0}^{n_1-1} \sum_{j=0}^{n_2-1} p_i^1 p_j^2 Z^{\varphi_\times(g_i^1, g_j^2)} \tag{7}$$

where $u_1(z)$ and $u_2(z)$ are the u-functions of the two random variables, $p_i^1$ and $p_j^2$ are the state probabilities, $g_i^1$ and $g_j^2$ are generic state values, and $\varphi_\times(g_i^1, g_j^2) = g_i^1 \times g_j^2$ is the structure function of state multiplication. It is noted that the composition operator $\otimes_\times$ corresponds to the multiplication-type structure function, which is a simple multiplication of two random variables and has not been formally specified previously.

For the solar generator, according to formula (4) $G^S = G^{MA} \times G^{IR}$, and based on the composition operator $\otimes_\times$ just introduced, the u-function of the solar generator power output $G^S$ can be written as:

$$\begin{aligned} u_S(z) &= u_{MA}^S(z) \otimes_\times u_{IR}(z) \\ &= \sum_{i=0}^{n_{IR}-1} \sum_{j=0}^{n_{MA}-1} p_i^{IR} p_j^{MA} Z^{g_i^{IR} \times g_j^{MA}} \\ &= \sum_{i=0}^{n_S-1} p_i^S Z^{g_i^S} \end{aligned} \tag{8}$$

where $n_S = n_{IR} \cdot n_{MA} - \delta_S$ is the total number of states of one solar generator ($\delta_S$ is the number of redundant states (i.e. states with the same amount of power output)). When computing the value of $n_S$, the collecting like-item technique is applied to reduce the number of items in the u-function (Li and Zuo 2007).

**2.2 Wind Turbine**

Similar to the solar, the wind turbine generation model consists of two parts: wind speed modeling and the turbine generation function. The Weibull distribution has been used to model the wind speed randomness (Boyle 2004):

$$f(v) = \frac{k}{c}\left(\frac{v}{c}\right)^{k-1} exp\left[-\left(\frac{v}{c}\right)^k\right] \tag{9}$$

where $k$ is the shape index, and $c$ is the scale index. When $k$ equals to 2, the probability density function is called Rayleigh density function.

Similar to solar irradiation, within a multi-state model the continuous wind speed distribution needs to be transformed into a discrete distribution. To this aim, the wind speed $v$ is divided into $n_{WS}$ states of equal size; the probability of the $i$th state can be obtained by:

$$Pr(v_i) = \int_{(i-1)\cdot \Delta v}^{i\cdot \Delta v} f(v)dv \tag{10}$$

where $\Delta v = v_{max}/n_{WS}$ is the step size, and $v_i$ is the expected value of wind speed in the $i$th state:

$$v_i = \frac{i\cdot \Delta v + (i-1)\cdot \Delta v}{2} \tag{11}$$

With the wind speed discretized into multiple states, the output of one wind turbine can be modeled by the following function (Hetzer et al. 2008):

$$P_W(v_i) = \begin{cases} 0 & v_i < v_{ci} \\ P_r^W \cdot \frac{(v_i - v_{ci})}{(v_r - v_{ci})} & v_{ci} \leq v_i < v_r \\ P_r^W & v_r \leq v_i < v_{co} \\ 0 & v_{co} \leq v_i \end{cases} \tag{12}$$

where $v_{ci}$ and $v_{co}$ are the cut-in, and cut-out wind speed respectively, $P_r^W$ is the rated power output, and $v_r$ is the rated wind speed.

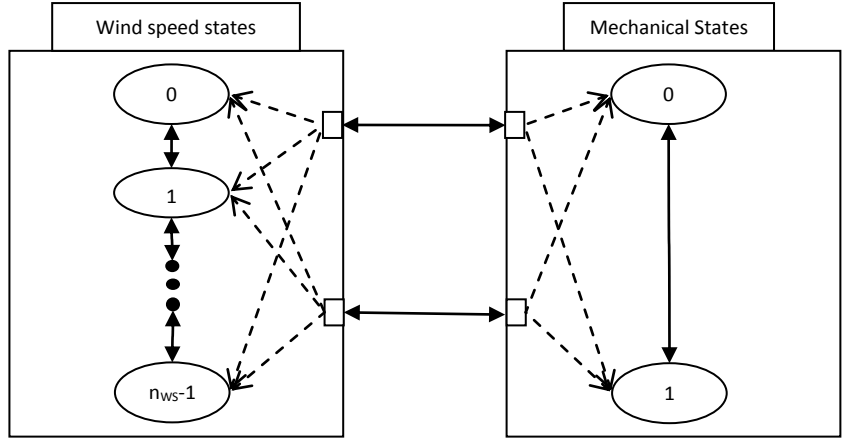

Fig 2. Wind speed states and mechanical states of the wind turbines

Similar to the solar generator, the wind turbine also contains two different sources of randomness – external and internal: the former is the wind speed and the latter is the mechanical degradation/failure/repair behavior (Guo et al. 2009, Arabian-Hoseynabadia et al. 2010, Nielsen and Sørensen, 2011). Generally, the wind turbine consists of three main subassemblies connected in series: generator, gearbox, and turbine rotor (Spinato et al. 2009). This implies that the failure of a single subassembly leads to the failure of the entire wind turbine.

Let $G^{WS}$ and $G^{MA}$ be the random variables representing the wind speed and mechanical condition, respectively. We assume that $G^{WS}$ and $G^{MA}$ are independent from each other and discretized into $n_{ws}$ and $n_{MA} = 2$ states, respectively. Let $\overrightarrow{Pr_{WS}} = \{p_0^{WS}, p_1^{WS}, \ldots, p_{n_{WS}-1}^{WS}\}$ and $\overrightarrow{Pr_{MA}} = \{p_0^{MA}, p_1^{MA}\}$ denote the state probability distributions of wind speed and mechanical state, respectively; $g_i^{IR}$ and $g_i^{MA} = \{0,1\}$ denote the state values of the wind speed and mechanical condition, respectively. The u-function of the wind speed state is:

$$u_{WS}(z) = \sum_{i=0}^{n_{WS}-1} p_i^{WS} z^{g_i^{WS}} \tag{13}$$

The u-function of the mechanical state is:

$$u_{MA}^W(z) = p_0^{MA} z^0 + p_1^{MA} z^1 \tag{14}$$

The overall u-function of the wind turbine can be obtained by:

$$u_W(z) = u_{WS}(z) \otimes_\times u_{MA}^W(z)$$
$$= \sum_{i=0}^{n_{WS}-1} \sum_{j=0}^{1} p_i^{WS} p_j^{MA} Z^{\varphi_\times(g_i^{WS}, g_j^{MA})} \quad (15)$$
$$= \sum_{i=0}^{n_W-1} p_i^W Z^{p_i^W}$$

where $\varphi_\times(\cdot)$ is the structure function, defined the same as for the solar generator, and $n_W = n_{WS} \cdot n_{MA} - \delta_W$ is the total number of states of the wind turbine ($\delta_W$ is the number of redundant states).

## 2.3 Electrical Vehicles

Electrical Vehicles (EV, or plug-in hybrid vehicles) can be important elements for distributed generation, with increasing expectation for the positive penetration of the system (Saber and Venayagamoorthy, 2011). An individual EV can be regarded as an electricity generator, a load or a storage, because it has a battery storage capable of charging, discharging and maintaining the power (Clement-Nyns et al. 2011). In our model, a group of $N_{EV}$ EVs is considered distributed on the system. Typically, these are modeled as moving like a single 'block group' and their power profiles are aggregated as a compound load, source or storage (Guille and Gross 2009, Clement-Nyns et al. 2011). The physical reasons for grouping EVs into one block are as follows: 1) the battery storage of one individual EV is too small to have influence on the power grid; 2) the majority of the vehicles follow a nearly stable daily usage schedule. A modeling reason for grouping EVs into one block is to avoid combinatorial explosion: if all EVs were considered separately, there would be $3^{N_{EV}}$ states of EV generation, load, and storage. In addition, it is assumed that all EVs are identical and each individual EV has two mechanical states: working or failed.

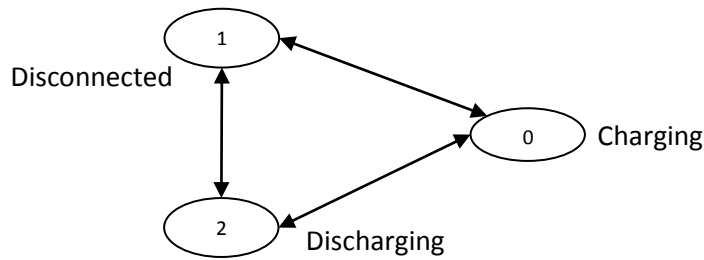

Fig 3. Operating states of a single EV

Similar to the solar generator modeling, the u-function of EV aggregation power output is obtained by multiplying the u-function of the power output of a single EV, representing its uncertain energy behavior, by the u-function of the mechanical condition of the EV aggregation, which indicates the number of functioning EVs. Let the variable $G^{OP}$ denote the random behavior of a single EV operation (represented by the multi-state model of Figure 3), with the states 0, 1, and 2 denoting the charging, disconnected, and discharging activities. Physically, $G^{OP}$ takes values from the set $\{g_0^{OP}, g_1^{OP}, g_2^{OP}\} = \{-P_v, 0, P_v\}$ in correspondence to the charging, disconnected, and discharging activities, respectively, and where $P_v$ is the absolute power output of one functioning EV. The probability vector of the set $\{-P_v, 0, P_v\}$ is denoted as $\overrightarrow{Pr_{OP}} = \{p_0^{OP}, p_1^{OP}, p_2^{OP}\}$, whose numerical values are $p_i^{OP} = t_i^{OP}/\sum_{i=0}^{2} t_i^{OP}$ where $t_i^{OP}$ is the time of residence of the EV in operation state $i$. The u-function of the operation of one EV can be written as:

$$u_{OP}(z) = p_0^{OP} Z^{g_0^{OP}} + p_1^{OP} Z^{g_1^{OP}} + p_2^{OP} Z^{g_2^{OP}} \tag{16}$$

Similar to the other renewable generators previously modeled, the EV aggregation is also assumed to be subject to mechanical degradation/failure/repair of its constituent EVs. Let $G^{MA}$ denote the mechanical condition of the EV aggregation, whose value represents the number of functioning EVs. The probability distribution of the mechanical state is denoted as $\overrightarrow{Pr_{MA}} = \{p_0^{MA}, p_1^{MA}, \ldots, p_{n_{MA}-1}^{MA}\}$. The u-function of the EV aggregation mechanical state is:

$$u_{MA}^{EV}(z) = \sum_{i=0}^{n_{MA}-1} p_i^{MA} Z^{g_i^{MA}} \tag{17}$$

The overall u-function of the EV aggregation can be obtained as:

$$\begin{aligned} u_{EV}(z) &= u_{OP}(z) \otimes_\times u_{MA}^{EV}(z) \\ &= \sum_{i=0}^{2} \sum_{j=0}^{n_{MA}-1} p_i^{OP} p_j^{MA} Z^{\varphi_\times(g_i^{OP}, g_j^{MA})} \\ &= \sum_{i=0}^{n_{EV}-1} p_i^{EV} Z^{g_i^{EV}} \end{aligned} \tag{18}$$

where $\varphi_\times(\cdot)$ is the property function, defined the same as for the solar generator and wind turbine, $g_i^{EV}$ is the power output (generation or consumption) by the EV aggregation (Saber and Venayagamoorthy, 2011), and $n_{EV} = n_{OP} \cdot n_{MA} - \delta_{EV}$ is the total number of states of the EV aggregation ($\delta_{EV}$ is the number of redundant states).

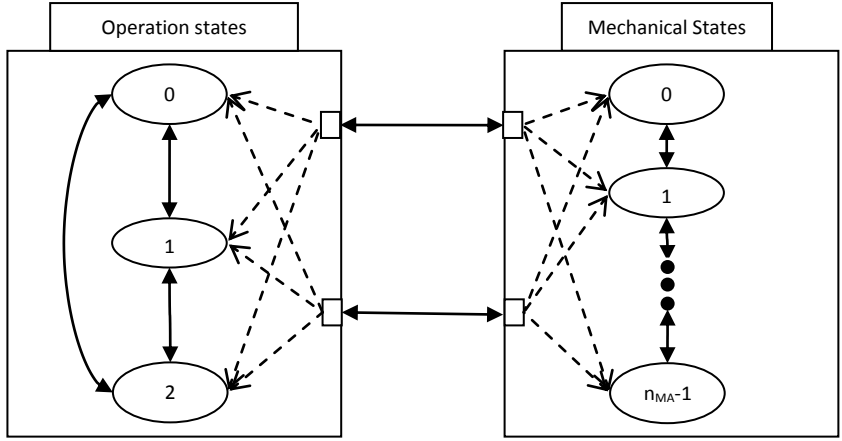

Fig 4. Operation states and mechanical states of the EV aggregation

## 2.4 Transformer

For the transformer, the randomness in its functional behavior is mainly due to its internal mechanical degradation/repair (Ding et al. 2011), which can be represented as a stochastic process of transitions among multiple states of degradation (Roos and Lindah 2004). Therefore, the UGF function for the transformer is defined as:

$$u_T(z) = \sum_{i=0}^{n_T-1} p_i^T z^{g_i^T} \tag{19}$$

Under Markovian assumption, the state probabilities can be obtained by solving the Markov stochastic transition model of Fig 5 with assigned transition rates (Massim et al. 2006).

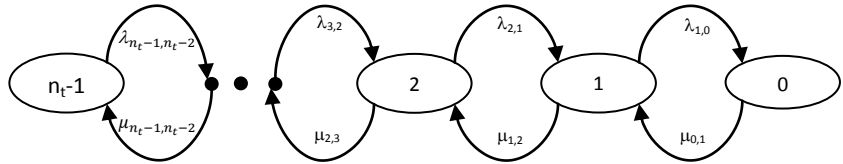

Fig 5. Mechanical states of a transformer

## 2.5 Load

In practice, the load values are typically recorded hourly on a specified time horizon (e.g. a year or a season). To model the dynamic behavior of loads, many multi-state models have been proposed ranging from a single load-aggregated representation up to more complex individual load modeling (Veliz et al. 2010). Load-aggregated models (Ding et al. 2011, Hegazy et al. 2003) resort to the load duration curve (LDC) model to reduce the number of load levels (Billinton and Allan 1996), and consider only one geological area pattern; differently, individual load modeling eventually resorts to a multilevel non-aggregate Markov model (Leite da Silva et al. 2000) which considers each hour as one state and includes the changing patterns in different areas. To keep the number of load states limited, we consider the LDC model which sorts all chronological load values into descending order of magnitude and divides the sorted load values into $n_L$ states (Fig. 6). The u-function of the multistate load model can be written as:

$$u_L(z) = \sum_{i=0}^{n_L-1} p_i^L z^{g_i^L} \tag{20}$$

where $l_i$ is the power consumption at state $i$ of the load.

## 3  Multi-State Model for the Distribution Network and Reliability Assessment

The following assumptions are made to combine the component models introduced in Section 2 to establish the multi-state model of the distributed generation system (Fig. 6):

(1) The distribution system under consideration is a local distribution network. It has a radial topology with one transformer and $m_L$ distribution nodes. In this topology, all components are actually connected in parallel, because they share a common feeder (transmission line) (Hegazy et al. 2003). The distribution nodes' load profiles are represented by a load-aggregate model.

(2) For the multiple $m_S$ solar and $m_W$ wind generators, the energy sources (i.e. solar irradiation and wind speed) are perfectly correlated, respectively. This assumption is reasonable for the local distribution network in a geographically close area and it can largely reduce the number of states of combined generators.

(3) For the wind and solar generators, the internal mechanical degradation/repair mechanism is mutually independent from each other. This is a common assumption in reliability modeling of hardware MSSs (Kuo and Prasad, 2000), especially when there is not enough data to describe the interdependence between components.

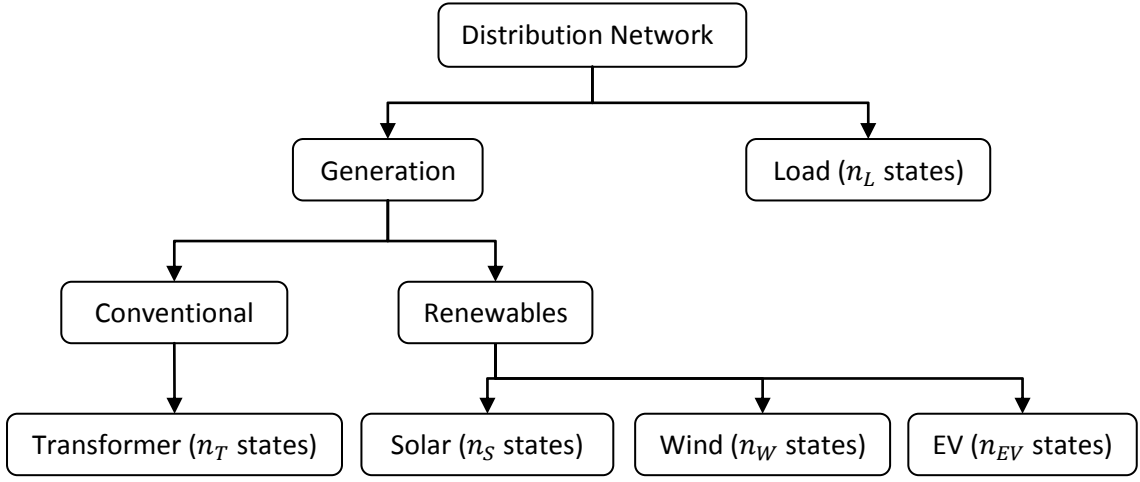

Fig 6. Hierarchy of the components of the distributed generation system

By assumptions (2) and (3), we obtain the u-function of the combined solar generators as follows:

$$u_S(z) = u_S^1(z) \otimes_+ \cdots \otimes_+ u_S^i(z) \otimes_+ \cdots \otimes_+ u_S^{m_S}(z)$$
$$= u_{IR}(z) \otimes_\times \left[ u_{MA}^{S_1}(z) \otimes_+ \cdots u_{MA}^{S_i}(z) \otimes_+ \cdots \otimes_\times u_{MA}^{S_{m_S}}(z) \right] \quad (21)$$

where $u_S^i(z)$ is the u-function of the $i$th solar generator, $u_{IR}(z)$ is the u-function of the solar irradiation for all solar generators, $u_{MA}^i(z)$ is the mechanical state u-function of the $i$th solar generator, and $m_S$ is the total number of solar generators. The UGF operator $\otimes_+$ between any two u-functions is defined as: $u(z) = u_1(z) \otimes_+ u_2(z) = \sum_{i=0}^{n_1-1} \sum_{j=0}^{n_2-1} p_i^1 p_j^2 Z^{\varphi_+(g_i^1, g_j^2)}$, where $\varphi_+(\cdot)$ is the property function representing the relationship $\varphi_+(g_i^1, g_j^2) = g_i^1 + g_j^2$.

Similarly, we can obtain the following UGF for the combined wind generator:

$$u_W(z) = u_W^1(z) \otimes_+ \cdots \otimes_+ u_W^i(z) \otimes_+ \cdots \otimes_+ u_W^{m_W}(z)$$
$$= u_{WS}(z) \otimes_\times \left[ u_{MA}^{W_1}(z) \otimes_+ \cdots u_{MA}^{W_i}(z) \otimes_+ \ldots \otimes_+ u_{MA}^{W_{m_W}}(z) \right] \quad (22)$$

where $u_W^i(z)$ is the u-function of the $i$th wind turbine, $u_{WS}(z)$ is the u-function of the wind speed for all wind turbines, $u_{MA}^i(z)$ is the mechanical state u-function for the $i$th wind turbine, and $m_W$ is the total number of wind turbines.

The u-function of all types of generators combined is:

$$u_G(z) = u_S(z) \otimes_+ u_W(z) \otimes_+ u_{EV}(z) \otimes_+ u_T(z)$$
$$= \sum_{i_S=0}^{n'_S-1} \sum_{i_W=0}^{n'_W-1} \sum_{i_{EV}=0}^{n_{EV}-1} \sum_{i_T=0}^{n_T-1} \left( p_i^S p_i^W p_i^{EV} p_i^T Z^{\varphi_+(g_i^S, g_i^W, g_i^{EV}, g_i^T)} \right) \quad (23)$$

where $n'_S = n_{IR}^S \cdot \prod_{i=1}^{m_S} n_{MA}^{S_i} - \delta_S$ (where $n_{IR}^S$ is the number of solar irradiation states, $n_{MA}^{S_i}$ is the number of mechanical states of the *i*th solar generator, and $\delta_S$ is the number of redundant states) and $n'_W = n_{WS}^W \cdot \prod_{i=1}^{m_W} n_{MA}^{W_i} - \delta_W$ (where $n_{WS}^W$ is the number of wind speed states, $n_{MA}^{W_i}$ is the number of mechanical states of the *i*th wind turbine, and $\delta_W$ is the number of redundant states) are the total numbers of states of the combined solar and wind generators, respectively. By further reducing the number of terms in (23), the u-function of the system generation takes the following expression:

$$u_G(z) = \sum_{i=0}^{n_G-1} p_i^G Z^{g_i^G} \quad (24)$$

where $n_G = n'_S \cdot n'_W \cdot n_{EV} \cdot n_T - \delta_G$ is the total number of energy states of the system generation and $\delta_G$ is the number of redundant states.

Given the assumption (1) above, the u-function of the aggregated load model has the form (20).

### 3.1 Reliability Assessment Indices

In MSS modeling, reliability (availability) is in general defined as the probability that the MSS lies in the states with capacity levels greater than or equal to the demand *W* (Zio, et al. 2007). The distributive operator Ψ has been proposed to obtain the reliability (availability) index $A(W)$ from the u-function of one MSS (Levitin et al. 1998):

$$\Psi(pZ^{g-W}) = \begin{cases} p, & \text{if } g \geq W \\ 0, & \text{if } g < W \end{cases} \quad (25)$$

$$A(W) = \Psi\left(\sum_{i=0}^{n-1} p_i Z^{g_i-W}\right) = \sum_{i=0}^{n-1} \Psi(p_i Z^{g_i-W}) \quad (26)$$

where *n* is the number of states of the MSS, and $g_i$ is the output of the MSS at state *i*.

For power systems, reliability (adequacy) is widely considered as a measure of the ability of the system power generation to meet the load demand (Billinton and Allan 1996, Rei and Schilling 2008, Levitin et al. 1998). Two common reliability assessment indices are loss of load expectation (*LOLE*) and expected energy not supplied (*EENS*) (Billinton and Allan 1996,

Rei and Schilling 2008). The former is the expected period during which the load demand is greater than available generation:

$$LOLE = \sum_{i=1}^{N_t} Pr(G_i < L_i) \tag{27}$$

where $i$ is the equally sized time step (e.g. hour or day), $N_t$ is the total number of time steps, $G_i$ is the total power generation available at time period $i$, $L_i$ is the load demand at time period $i$, $Pr(G_i < L_i)$ is the probability that the load demand exceeds the available power generation at time period $i$. The latter is the expectation of the energy that the system is not able to supply:

$$EENS = \sum_{i=1}^{N_t} Pr(G_i < L_i) \times (L_i - G_i) \tag{28}$$

where $L_i - G_i$ is the energy that the system is not able to supply at time step $i$. Typically, these two indices are computed through Monte Carlo simulation (Billinton and Allan 1996). In this study, we utilize the distributive operator $\Psi$ to compute LOLE and EENS from the u-functions of system generation in (25) and load demand in (20). The LOLE index is written as:

$$\begin{aligned} LOLE &= N_t \cdot \Psi\left(\sum_{i=0}^{n_G-1} \sum_{j=0}^{n_L-1} p_j^L p_i^G z^{g_j^L - g_i^G}\right) \\ &= N_t \cdot \sum_{i=0}^{n_G-1} \sum_{j=0}^{n_L-1} \Psi(p_j^L p_i^G z^{g_j^L - g_i^G}) \end{aligned} \tag{29}$$

Similarly, *EENS* is written as:

$$EENS = N_t \cdot \sum_{i=0}^{n_G-1} \sum_{j=0}^{n_L-1} (g_j^L - g_i^G) \Psi(p_j^L p_i^G z^{g_j^L - g_i^G}) \tag{30}$$

## 4 Case Study

The system used as case study is modified from the IEEE 34 node distribution test feeder (Kersting 1991), and is a radial distribution network downscaled to 4.16 kV via the in-line transformer. In this network, the rated power of the transformer is 5000 kW (Kersting 1991). The common assumption made for the transformer, solar generator, wind generator and EV aggregation is that the random process of internal mechanical degradation/repair can be modeled as a Markov process of stochastic transitions between the two states of working and failed. This brings no loss of generality with respect to the UGF modeling approach.

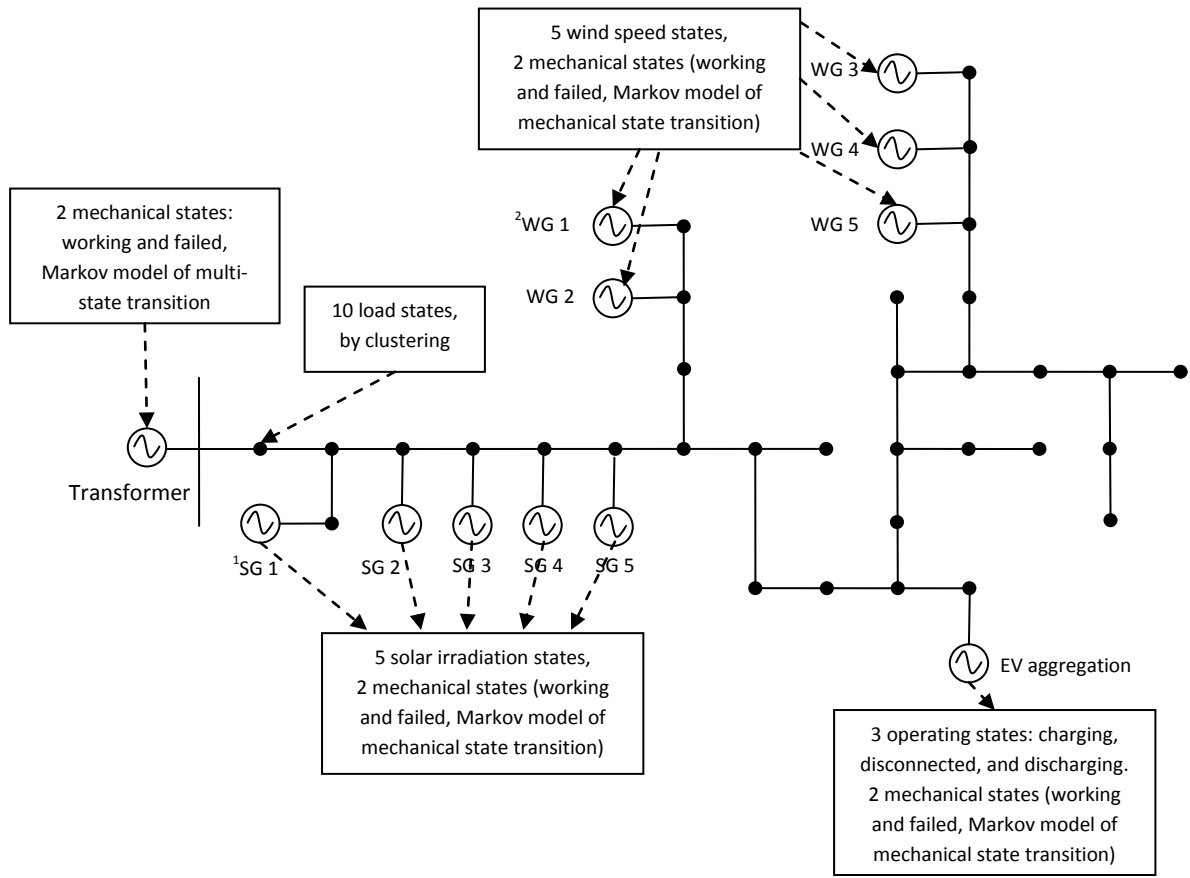

Fig 7. IEEE 34 nodes distribution test feeder modified for distributed generation

[1] SG: solar generator, [2] WG: wind generator

The ratio of renewable energy to conventional energy is 25% (Ackermann et al. 2001). Within the renewable energy, wind, solar, and EV occupy a share of 60%, 30% and 10%, respectively. Therefore, the rated power outputs of wind, solar and EV are 750 kW, 375 kW, and 125 kW, respectively. The ratio between wind energy and solar energy represents the recent situations of some European countries such as Czech Republic and Austria (Eurostat, 2009). The EV aggregation is taken as a share of 10% because it is a relatively new technology still undergoing development. The renewable generation source consists of five identical solar generators (each one containing 1000 solar modules which is with 75 W rated power), five identical wind turbines (each with 150 KW rated power), and an EV aggregation containing 25 EVs (each with 5KW rated power).

The reliability block diagram (RBD) representation of the distributed generation system is given in Figure 8.

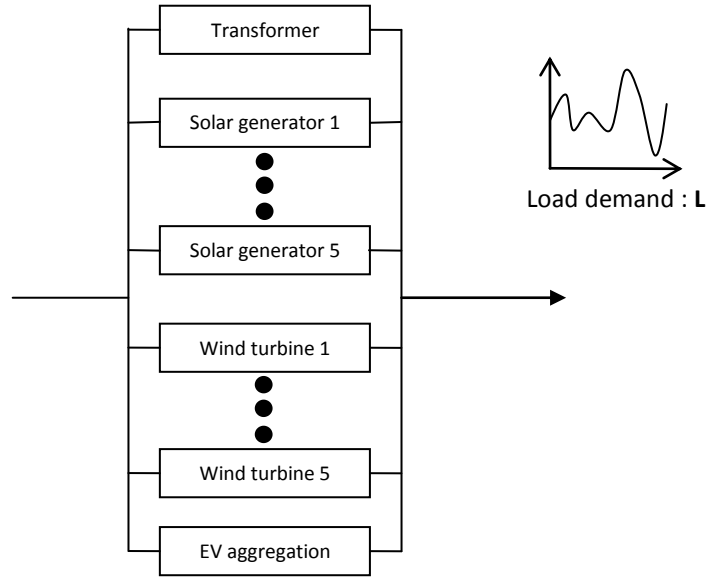

Fig 8. Reliability block diagram of the distributed generation system

It is shown by Figure 8 that all the power generation sources are modeled as connected in a parallel logic structure with respect to the function of providing the power to satisfy the load demand. The transformer is represented by a Markov model with two states: working and failed. The failure and repair rates are 0.0004/yr and 0.013/yr (Roos and Lindah 2004), respectively. By solving the Markov model, the steady probabilities of the working and failure states are 0.97 and 0.03, respectively. The UGF for the transformer is then:

$$u_T(z) = 0.97Z^{5000} + 0.03Z^0$$

As described in Section 2.1, the model of solar generator can handle the failure of any portion of the solar modules and the uncertain state of solar irradiance. Due to the limitation of the data available, two mechanical states (e.g. all solar modules are working or all are failed) are considered for the numerical example, with failure and repair rates set to 0.0005/hr and 0.013/hr, respectively (Karki and Billinton 2001). After solving the Markov model, the steady probabilities of working and failed states occupancy are 0.96 and 0.04, respectively.

The parameters of the Beta distribution of solar irradiation have been estimated by fitting the average daily solar irradiation data taken from Mohamed and Koivo (2010). The resulting distribution is divided into 5 equally sized states with different interval sizes and corresponding probabilities, to search for the optimal number of states (Billinton and Gao 2008). The state probabilities are computed by eqs. (2-3). For each state, the expected solar irradiation value is substituted into the generation function (4) to obtain the expected power output of a single solar module. Table 1 shows the information of the 5-state division.

Table 1. five-state solar model of a single solar module

| State No. | Solar irradiation (kW/m2) | Probability | Power output (kW) |
|---|---|---|---|
| 1 | 0.1 | 0.59 | 0.00825 |
| 2 | 0.3 | 0.13 | 0.024 |
| 3 | 0.5 | 0.10 | 0.0405 |
| 4 | 0.7 | 0.08 | 0.05625 |
| 5 | 0.9 | 0.10 | 0.072 |

In the case of 5-state division, the UGF for the compound of all five solar generators is:

$$u_S(z) = u_{IR}(z) \otimes_\times [u_{MA}^1(z) \otimes_+ u_{MA}^2(z) \otimes_+ u_{MA}^3(z) \otimes_+ u_{MA}^4(z) \otimes_+ u_{MA}^5(z)]$$

$$= (0.59Z^{0.00825} + 0.13Z^{0.024} + 0.1Z^{0.0405} + 0.08Z^{0.05625} + 0.1Z^{0.072}) \otimes_\times [(0.96Z^{1000} + 0.04Z^0) \otimes_+ \dots (0.96Z^{1000} + 0.04Z^0)]$$

$$= (0.59Z^{0.00825} + 0.13Z^{0.024} + 0.1Z^{0.0405} + 0.08Z^{0.05625} + 0.1Z^{0.072}) \otimes_\times (1.02 \times 10^{-7}Z^0 + 1.23 \times 10^{-5}Z^{1000} + 5.90 \times 10^{-4}Z^{2000} + 0.0142Z^{3000} + 0.170Z^{4000} + 0.815Z^{5000})$$

$$= \underbrace{1.02 \times 10^{-7}Z^0 + 7.25 \times 10^{-6}Z^{8.09} + \dots + 0.0815Z^{360}}_{26\ items}$$

The individual wind turbine is modeled with two mechanical states: complete failure and perfect functioning. The failure and repair rates are set to 0.0005/hr and 0.013/hr, respectively (Karki and Billinton, 2001). The Rayleigh distribution of wind speed (obtained by fitting the average daily wind speed data taken from Mohamed and Koivo (2010)) has been divided into 5 equally sized states, with the associated probabilities and power outputs (see Table 2).

Table 2. five-state wind model of a single wind turbine

| State No. | Wind Speed (km/hr) | Probability | Power output (kW) |
|---|---|---|---|
| 1 | 4 | 0.39 | 2.85 |
| 2 | 12 | 0.47 | 36 |
| 3 | 20 | 0.12 | 69 |
| 4 | 28 | 0.011 | 100.5 |
| 5 | 36 | 0.003 | 133.5 |

Similarly, the UGF of the combined five wind generators is:

$$u_W(z) = u_{WS}(z) \otimes_\times [u_{MA}^1(z) \otimes_+ u_{MA}^2(z) \otimes_+ u_{MA}^3(z) \otimes_+ u_{MA}^4(z) \otimes_+ u_{MA}^5(z)]$$

$$= (0.39Z^{2.85} + 0.47Z^{36} + 0.12Z^{69} + 0.011Z^{100.5}$$
$$+ 0.003Z^{133.5}) \otimes_\times [(0.96Z^1 + 0.04Z^0) \otimes_+ \ldots (0.96Z^1 + 0.04Z^0)]$$

$$= (0.39Z^{2.85} + 0.47Z^{36} + 0.12Z^{69} + 0.011Z^{100.5} + 0.003Z^{133.5}) \otimes_\times (1.02$$
$$\times 10^{-7}Z^0 + 1.23 \times 10^{-5}Z^1 + 5.90 \times 10^{-4}Z^2 + 0.0142Z^3 + 0.170Z^4$$
$$+ 0.815Z^5)$$

$$= \underbrace{1.02 \times 10^{-7}Z^0 + 4.79 \times 10^{-6}Z^{2.87} + \cdots + 0.00245Z^{668}}_{26\ items}$$

For the EV aggregation, the operation state probabilities are obtained by using the daily charging profile (Clement-Nyns et al. 2011): $\overrightarrow{Pr_{OP}} = \left\{\frac{3}{24}, \frac{20}{24}, \frac{1}{24}\right\} = \{0.13, 0.83, 0.04\}$. The rated power of a single EV is 5kW, which corresponds to a random variable of EV aggregation $G^{OP}$ with values in the set $\{-5, 0, 5\}$. The mechanical state probabilities are obtained by using the failure and repair rates 0.0013/hr and 0.12/hr, respectively: $\overrightarrow{Pr_{MA}} = \{0.99, 0.01\}$. Then the overall UGF becomes:

$$u_{EV}(z) = u_{OP}(z) \otimes_\times u_{MA}^{EV}(z) = (0.13Z^{-5} + 0.83Z^0 + 0.04Z^5) \otimes_\times (0.01Z^0 + 0.99Z^{25})$$
$$\approx 0.13Z^{-125} + 0.83Z^0 + 0.04Z^{125}$$

By combining the UGFs of all generators, we obtain the composite generation u-function for the 5-state division of both wind speed and solar irradiation:

$$u_G(z) = u_S(z) \otimes_+ u_W(z) \otimes_+ u_{EV}(z) \otimes_+ u_T(z)$$
$$= \left(\underbrace{1.02 \times 10^{-7}Z^0 + 7.25 \times 10^{-6}Z^{8.09} + \cdots + 0.0815Z^{360}}_{26\ items}\right)$$
$$\otimes_+ \left(\underbrace{1.02 \times 10^{-7}Z^0 + 4.79 \times 10^{-6}Z^{2.87} + \cdots + 0.00245Z^{668}}_{26\ items}\right)$$
$$\otimes_+ (0.13Z^{-125} + 0.83Z^0 + 0.04Z^{125}) \otimes_+ (0.03Z^0 + 0.97Z^{5000})$$
$$= \left(\underbrace{4.32 \times 10^{-7}Z^{-94.5} + 4.79 \times 10^{-6}Z^{-92.2} + \cdots + 7.74 \times 10^{-6}Z^{6153}}_{1438\ items}\right)$$

The multi-state load model is built on the 8736 hourly load values of the IEEE-reliability testing system (IEEE-RTS), with a peak load of 5500 kW and minimal load of 1863.5 kW.

This value satisfies the ratio of average peak load to average transformer power output in Hegazy et al. (2003). The load values are grouped into ten equally-sized intervals in the range (1863.5, 5500) kW for a reasonable trade-off between modeling accuracy and evaluation efficiency (Singh and Lago-Gonzales 1989). The probability for each load interval/state is defined as the ratio of the number of load values inside the interval to the total number of load values. For example, the state probability of the first interval/state is $\frac{384}{8736} = 0.44$. The state value of each interval/state is the average of the lower and upper bounds of the interval. For example, the performance value of the first interval/state is $\frac{1863.5+2227.2}{2} \approx 2045$. After the load value clustering and the state probability computation, we can obtain the final u-function for the load:

$$u_L(z) = 0.044Z^{2045} + 0.137Z^{2408} + 0.174Z^{2773} + 0.131Z^{3136} + 0.161Z^{3500}$$
$$+ 0.124Z^{3864} + 0.110Z^{4227} + 0.088Z^{4591} + 0.029Z^{4955} + 0.004Z^{5318}$$

Based on the u-functions of system generation and load, the reliability indices for the DG system are computed:

$$LOLE = 8736 \cdot \sum_{i=0}^{1437} \sum_{j=0}^{9} \Psi(p_j^L p_i^G Z^{g_j^L - g_i^G}) = 259.52 \text{ hr/yr}$$

$$EENS = 8736 \cdot \sum_{i=0}^{1437} \sum_{j=0}^{9} (g_j^L - g_i^G) \Psi(p_j^L p_i^G Z^{g_j^L - g_i^G}) = 822.45 \text{ MWhr/yr}$$

## 5 Discussion and Conclusion

To the knowledge of the authors, this study is a first in proposing a UGF-based multi-state analytical model for the reliability assessment of a distributed generation system with renewable energy sources. Multi-state sub-models are built for each element in the distribution network, including solar generator, wind generator, transformer, electrical vehicles and load. The UGF is used to mathematically represent the multi-state elements and combine their states through a formally introduced composition operator, to obtain the final system model which allows computing the reliability indices in a straightforward manner. An illustration of the method has been provided, with respect to a modified IEEE 34 node test feeder distribution network.

For small-scale power grids, analytical modeling can be efficient, accurate, and serve the purpose of providing reference results. However, some limitations exist in the study presented, which origin from some of the assumptions underlying the models. For example, the dependence between the energy variables and mechanical states of the solar and wind generators respectively, are not considered. In practice, over-intensity of irradiation and over-speed of wind may cause damage to the internal components of the generators and lead to mechanical degradation and failures. In addition, transmission lines are neglected whereas their failures can result in 'islanding' of the downstream area of the network.

Besides the above modeling issues, the computational effort needed to solve the model may also be critical, depending on the number of components and states considered. UGF operation is essentially a convolution computation: development of more efficient algorithms can be considered from two directions: 1) combining states to reduce the number of states by clustering techniques, 2) fast convolution algorithms (such as fast Fourier transform (FFT)).

**Acknowledgement**

We want to show our appreciation to the referees for carefully reviewing the paper and providing valuable comments which have helped substantially improving the paper.